# Raman and IR Signatures of $[Mo_3S_4]^{4+}$ and $[Mo_3S_{13}]^{2-}$ Molybdenum Sulphide Molecular Catalysts for Solar Hydrogen Evolution

**Pardis Adams[1], Jan Bühler[1], Angel Labordet Alvarez[2,3], D. David Tilley[1], and Mirjana Dimitrievska[2,*]**

[1] Department of Chemistry, University of Zurich, Zurich, Switzerland.
[2] Nanomaterials Spectroscopy and Imaging, Transport at Nanoscale Interfaces, Swiss Federal Laboratories for Materials Science and Technology (Empa), Dubendorf, Switzerland.
[3] Department of Physics and Swiss Nanoscience Institute, University of Basel, Basel, Switzerland

*Author to whom any correspondence should be addressed.

E-mail: Mirjana.Dimitrievska@empa.ch

## Abstract

Molybdenum sulphide clusters $[Mo_3S_4]^{4+}$ and $[Mo_3S_{13}]^{2-}$ are key molecular models for Mo–S-based catalysts and promising candidates for energy conversion, yet their vibrational properties remain underexplored. This study presents a comprehensive Raman and infrared (IR) spectroscopic analysis of both clusters, supported by density functional theory (DFT) calculations. High-quality crystalline samples were prepared and characterised via scanning electron microscopy (SEM) and energy-dispersive X-ray spectroscopy (EDX). Raman spectra, acquired with 488 and 532 nm lasers and fitted with Lorentzian curves, were assigned through direct comparison with DFT-predicted vibrational modes. For $[Mo_3S_4]^{4+}$, bands at ~200, 350, and 450 cm$^{-1}$ correspond to Mo–S–Mo bending, Mo–S stretching, and terminal S vibrations. $[Mo_3S_{13}]^{2-}$ shows bands in two regions: 100–400 cm$^{-1}$ (Mo–S/S–S modes) and 450–550 cm$^{-1}$ (terminal S–S stretching). IR spectra provided additional vibrational features, providing a more complete fingerprint for each cluster. Raman spectroscopy proved more sensitive than X-ray diffraction (XRD) for identifying these when supported on other types of materials such as $Sb_2Se_3$. This work offers a detailed vibrational reference for $[Mo_3S_4]^{4+}$ and $[Mo_3S_{13}]^{2-}$, establishing Raman and IR spectroscopy as robust techniques for characterising Mo–S molecular clusters in both fundamental and applied research.

## 1. Introduction

Molybdenum sulphide ($Mo_xS_y$) clusters have emerged as a promising class of catalysts for a wide range of chemical transformations, particularly in sustainable energy and green chemistry.[1,2] Among these, molecular clusters such as $[Mo_3S_4]^{4+}$ and $[Mo_3S_{13}]^{2-}$ have attracted increasing interest due to their well-defined structures, tunable reactivity, and capacity to mimic the active sites of heterogeneous molybdenum sulphide catalysts.[3,4] Molybdenum disulphide ($MoS_2$), a layered transition metal dichalcogenide, has long been recognised for its catalytic activity in hydrodesulphurisation (HDS)[5,6] and, more recently, for the hydrogen evolution reaction (HER)[7–10]. However, its catalytic performance is primarily governed by edge sites, which are difficult to control and characterise in bulk materials.[11–14] On the other hand, molecular clusters such as $[Mo_3S_4]^{4+}$ and $[Mo_3S_{13}]^{2-}$ provide a structurally precise and atomically defined alternative with tuneable redox properties. These clusters serve as a model system for understanding active sites in heterogeneous catalysts, allowing systematic investigation of structure-activity relationships and facilitating mechanistic insights at the molecular level.[15–17]

The $[Mo_3S_4]^{4+}$ cluster, often seen as a core motif in larger molybdenum sulphide clusters and $MoS_2$ edge sites,[18,19] comprises a $Mo_3$ triangle with an apical sulphur atom and bridging sulphide ligands (Fig. 1a). This cluster has been explored for its ability to mediate multi-electron transformations and is particularly interesting as a structural analogue to biological molybdenum cofactors found in enzymes such as nitrogenases and sulphite oxidases.[2,20–23] Similarly, the $[Mo_3S_{13}]^{2-}$ cluster features a triangular molybdenum core bridged by sulphur atoms and capped with terminal disulphide ligands (Fig. 1b). This structure exhibits rich redox chemistry and accessible active sites that can catalyse electron and proton transfer cycles.[24,25] Studies have shown that $[Mo_3S_{13}]^{2-}$ can catalyse HER with performance metrics approaching those of bulk $MoS_2$ while providing a molecular platform amenable to fine-tuning through ligand substitution or cluster derivatisation.[26–32] Collectively, $[Mo_3S_4]^{4+}$ and $[Mo_3S_{13}]^{2-}$ represent key models for understanding catalytic processes at Mo–S active sites.[2,33,34] Their modularity enables a bottom-up approach to catalyst design, bridging the gap between molecular and solid-state catalysis.[3,24,35]

Continued exploration and comprehensive characterisation of these clusters hold promise for advancing the rational development of new catalysts for hydrogen generation, $CO_2$ reduction, and other sustainable chemical processes.[36–38] Raman spectroscopy has proven to be an indispensable technique for characterising these molecular clusters.[39–47] As a vibrational analysis technique, Raman spectroscopy offers detailed fingerprints for identifying structural motifs and ligand environments. It also functions as an effective probe for assessing compositional integrity and observing transformations under operando conditions.[48]

In particular, for clusters like $[Mo_3S_4]^{4+}$ and $[Mo_3S_{13}]^{2-}$, where sulphur coordination and bonding are central to the function, Raman spectroscopy is a powerful tool because it provides detailed insight into local bonding environments, including metal–sulphur and sulphur–sulphur interactions. It is sensitive to changes in coordination, oxidation state, and symmetry, enabling the identification of structural motifs, defects, and phase variations that are often undetectable by techniques like X-ray diffraction.[49] Raman can also monitor cluster stability and transformations under different conditions.[50] Raman also complements infrared spectroscopy by detecting IR-inactive modes, such as terminal disulphide stretches, enabling a more complete vibrational fingerprint. Its non-destructive nature, minimal sample requirements, and ability to track changes in geometry or oxidation state make it especially valuable for studying Mo–S clusters in both fundamental and applied contexts.

To date, Raman data for $[Mo_3S_4]^{4+}$ and $[Mo_3S_{13}]^{2-}$ have been reported in a few studies, but typically in a partial or application-specific context.[51] For example, Raman spectra of $[Mo_3S_{13}]^{2-}$ have been observed during electrocatalytic operation and thin-film preparation, often showing characteristic bands in the 200–550 $cm^{-1}$ range associated with Mo–S and S–S vibrations.[25] Similarly, $[Mo_3S_4]^{4+}$ species have been investigated as precursors or intermediates in sulphide synthesis, but their complete vibrational assignments are rarely presented or compared to density functional theory (DFT) calculations.[52] This lack of a well-established vibrational reference dataset poses challenges for researchers aiming to synthesise, characterise, or detect these clusters reliably, especially in complex or hybrid materials environments.

While previous studies have reported partial or application-specific Raman observations for these clusters, none have provided complete mode assignments across the whole spectral range, nor directly benchmarked them fully against theory. This study addresses this need by providing a complete vibrational characterisation of the $[Mo_3S_4]^{4+}$ and $[Mo_3S_{13}]^{2-}$ clusters using a combined experimental and computational approach. This begins by preparing high-quality crystalline samples of each cluster and verifying their morphology, composition, and stoichiometry using scanning electron microscopy (SEM) and energy-dispersive X-ray spectroscopy (EDX). Raman spectroscopic measurements are performed using two excitation wavelengths (532 nm and 488 nm), allowing us to probe resonance effects and improve mode detection across a broad spectral range. These measurements are directly compared with DFT predictions, enabling the assignment of all experimentally observed Raman modes. In parallel, the calculated infrared (IR) vibrational spectra are reported, and the corresponding IR-active modes are identified for both clusters. By presenting both Raman and IR signatures along with full DFT-supported assignments, a comprehensive vibrational fingerprint for $[Mo_3S_4]^{4+}$ and $[Mo_3S_{13}]^{2-}$ is established. This approach supports the precise identification of these clusters and opens new avenues for extending their characterisation via IR spectroscopy, particularly in environments where Raman scattering may be limited. This dual experimental–computational framework establishes a robust structural "fingerprint" for Mo–S molecular catalysts, filling a critical gap that has hindered reliable characterisation in catalysis and energy conversion research.

## 2. Results and Discussion

### *2.1 Structural Overview of $[Mo_3S_4]^{4+}$ and $[Mo_3S_{13}]^{2-}$ Clusters*

The molecular structures of $[Mo_3S_4]^{4+}$ and $[Mo_3S_{13}]^{2-}$ are shown in Fig. 1, highlighting the differences in nuclearity, sulphur coordination, and symmetry that underlie their spectroscopic behaviour and chemical functionality. The $[Mo_3S_4]^{4+}$ cluster in Fig. 1a features a triangular arrangement of three Mo atoms bridged by a tetrahedral sulphur ligand environment. The cluster includes three $\mu_2$-sulphide ligands, sulphide ions ($S^{2-}$) that bridge two Mo atoms along the triangle's edges, and a single $\mu_3$-sulphide that caps one face of the triangle by coordinating all three Mo atoms. Crystallographic data[51,53,54] show that the Mo–Mo distances range from 2.801 to 2.82 Å, indicating significant metal-metal bonding. The Mo–S bond lengths vary depending on the coordination environment: Mo–($\mu_2$-S) bonds are 2.28–2.30 Å, while Mo–($\mu_3$-S) bonds are slightly longer, ranging from 2.32–2.33 Å, due to the higher coordination number of the apical sulphur atom. Each Mo centre adopts a distorted tetrahedral geometry, coordinated by two bridging sulphides, one apical $\mu_3$-sulphide, and one or more metal-metal interactions. This compact, symmetric configuration reflects structural motifs found at catalytically active edge sites in $MoS_2$ and provides a molecular model for sulphur-mediated multi-electron processes.[3,55] In contrast, the $[Mo_3S_{13}]^{2-}$ cluster, shown in Fig. 1b, exhibits a more open and sulphur-rich architecture. It contains a similar triangular $Mo_3$ core but is surrounded by thirteen sulphur atoms, comprising both terminal and bridging species. Notably, the cluster includes six terminal disulphide ($S_2^{2-}$) ligands, each coordinating to a Mo centre. These disulphides show Mo–S bond lengths in the range of 2.38–2.45 Å, which are longer than those of bridging sulphides, reflecting weaker Mo–S interactions. The remaining sulphur atoms form a combination of $\mu_2$- and $\mu_3$-bridging ligands, with Mo–S distances between 2.30–2.36 Å for $\mu_2$-S and up to 2.40 Å for $\mu_3$-S bridges.

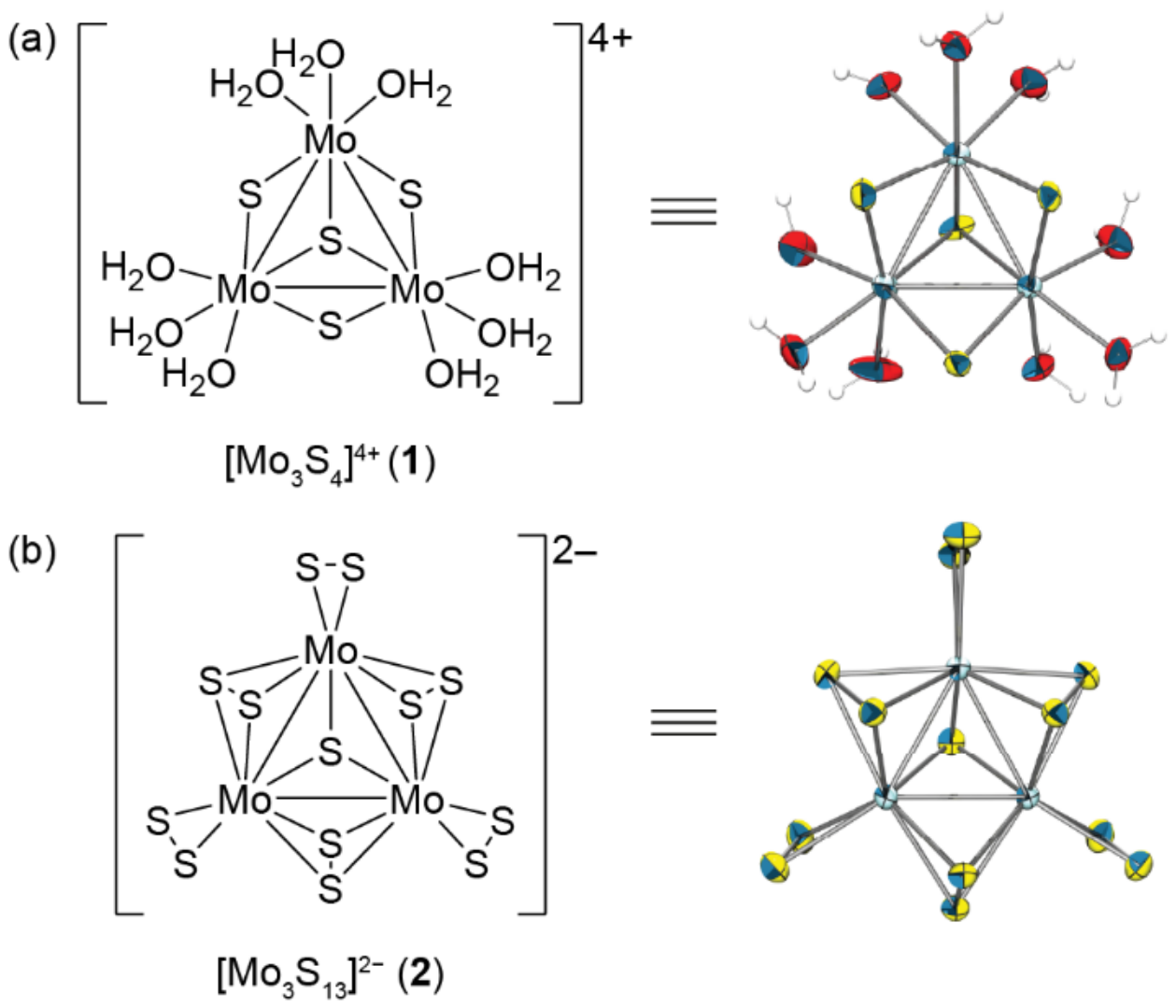

**Figure 1.** Molecular structure and ellipsoid displacement plots of (a) $[Mo_3S_4]^{4+}$ and (b) $[Mo_3S_{13}]^{2-}$. Ellipsoids represent 50% probability. Solvent molecules and counterions ($Cl^-$ for (b) and $NH^{4+}$ for (c)) are omitted for clarity. Figure modified from [51].

### *2.2 Morphological and Compositional Analysis*

To verify the crystallinity, morphology, and elemental composition of the prepared clusters, scanning electron microscopy (SEM) and energy-dispersive X-ray spectroscopy (EDX) was performed on isolated crystals of $[Mo_3S_4]^{4+}$ and $[Mo_3S_{13}]^{2-}$. These analyses are crucial to confirm that the synthetic procedures yield phase-pure, structurally intact materials prior to spectroscopic characterisation. Fig. 2 shows SEM images and corresponding compositional maps of Sn, S and Mo measured with EDX. The presence of tin (Sn) in the spectra is attributed to the use of fluorine-doped doped-$SnO_2$-coated (FTO) substrates during the measurements. SEM imaging revealed well-defined microcrystals with distinct morphological features reflective of their underlying structural and packing motifs. As observed in Fig. 2a, crystals of $[Mo_3S_4]^{4+}$ appear as uniform, prismatic blocks, consistent with compact molecular packing driven by the cluster's relatively symmetric, triangular core structure. In contrast, Fig. 2e shows that $[Mo_3S_{13}]^{2-}$ forms more irregular, platelet-like crystallites, which may be attributed to its higher sulphur content, extended ligand shell, and more open coordination environment. These morphological differences align with the divergent molecular architectures of the two clusters and suggest that subtle variations in cluster geometry influence their nucleation and growth behaviour during crystallisation. EDX analysis in Fig. 2b-d and Fig. 2f-h confirmed the elemental composition of both clusters, with clearly resolved molybdenum and sulphur signals. As determined by EDX, shown in Table S1, the Mo:S atomic ratio for $[Mo_3S_4]^{4+}$ was approximately 0.83±0.10, in good agreement with the theoretical value of 0.75. For $[Mo_3S_{13}]^{2-}$, the Mo:S ratio was found to be 0.31±0.10, compared to the ideal 0.23 expected from the cluster stoichiometry. Both values are in agreement with the theoretically expected stoichiometry within the error limit. The underrepresentation of sulphur may be attributed to the limited sensitivity of EDX to light elements, particularly when present in the terminal disulphide ($S_2^{2-}$) form. Additionally, the close energy spacing between the Mo Mα and S Kα emission lines (~2.29 and ~2.31 keV, respectively) can lead to peak overlap, introducing further uncertainty in independent quantification. Altogether, the single crystal XRD, SEM and EDX results confirm both clusters' morphology and elemental composition and provide a solid basis for the following vibrational spectroscopic analysis.

## *2.2 Raman Spectroscopy and Vibrational Mode Assignment*

In layered Mo–S materials such as $MoS_2$, the Raman spectrum primarily reflects collective lattice vibrations of the extended crystal, with the dominant $E_g$ and $A_{1g}$ modes being only indirectly influenced by synthesis and processing. Changes in preparation route, such as chemical vapour deposition, spin-coating, or chemical exfoliation, tend to affect the fingerprint through alterations in thickness, polytype distribution (2H vs. 1T), defect density, or disorder, sometimes activating normally forbidden modes due to relaxed selection rules. In contrast, molecular Mo–S catalysts, which lack an extended basal plane and consist instead of discrete Mo–S clusters, exhibit Raman features arising from well-defined local vibrations such as terminal and bridging Mo–S stretches and S–S bonds. In these systems, synthesis and processing directly modulate the immediate chemical environment of catalytically active sites, leading to predictable shifts in sharp, well-assigned vibrational bands. As a result, while Raman spectroscopy in layered films often acts as an indirect probe of catalytic structure, in molecular Mo–S systems it can serve as a direct and sensitive reporter of active-site chemistry.[56–58]

Before analysing the vibrational spectra of $[Mo_3S_4]^{4+}$ and $[Mo_3S_{13}]^{2-}$, it is essential to consider the symmetry of the molecular structures, as symmetry dictates the selection rules and classification of normal modes into irreducible representations. In an idealised, isolated form, the $[Mo_3S_4]^{4+}$ cluster possesses a nearly equilateral triangular $Mo_3$ core capped by a $\mu_3$-sulphide and bridged by three $\mu_2$-sulphides. This geometry approximates $C_{3V}$ point group symmetry, resulting in a set of Raman-active $A_1$ and E modes and IR-active E modes.[59–61] $[Mo_3S_{13}]^{2-}$ cluster is more complex due to the presence of multiple inequivalent disulphide ligands and asymmetric bridging sulphurs. In this case, the overall structure lacks symmetry elements beyond identity and is best described by the $C_1$ point group. In this case, all vibrational modes are formally active in both Raman and infrared spectroscopy, and no degeneracy or selection rules apply. Furthermore, even in the case of $[Mo_3S_4]^{4+}$, deviations from idealised symmetry due to crystal packing effects, solvation, or thermal distortions can reduce the effective symmetry below $C_{3v}$. To treat both clusters consistently and to enable detailed comparison between experiment and theory, vibrational mode analysis and DFT calculations were performed assuming no symmetry constraints ($C_1$). This approach ensures that all vibrational degrees of freedom are included, and that subtle distortions and mode couplings are captured in the calculations and, therefore, in the calculated vibrational mode properties. Following the symmetry considerations, Fig. 3 presents the experimental and DFT-calculated Raman spectra of the $[Mo_3S_4]^{4+}$ and $[Mo_3S_{13}]^{2-}$ clusters. The upper panels show the measured spectra obtained using 532 nm and 488 nm excitation wavelengths, highlighting all Raman-active vibrational modes observed under ambient conditions. Both excitation sources yielded consistent results, with no significant resonance enhancement or wavelength-dependent shifts, confirming the robustness of the spectral features across these two excitation energies. The corresponding DFT-calculated Raman spectra are displayed in black alongside the experimental ones in Fig. 3e. These theoretical Raman spectra were obtained from DFT calculations performed without symmetry constraints ($C_1$ point group) to fully account for all Raman-active normal modes, including those arising from subtle structural distortions. The calculated spectra were normalised to align with the experimental data and allow direct mode-by-mode comparison. Beneath the spectra, Fig. 3 also displays vertical bars marking the positions of DFT-predicted Raman-active modes and experimentally extracted peak frequencies. Experimental peak frequencies were determined by a rigorous Lorentzian deconvolution procedure, as described in detail by Dimitrievska et al. (2023,2024),[62,63] allowing accurate separation of overlapping features and assignment of vibrational modes with improved accuracy. This method enabled the identification of weak and partially overlapping peaks that would otherwise be obscured in the spectra, particularly in the congested mid and high-frequency regions of the $[Mo_3S_{13}]^{2-}$ spectrum. The experimental and computational analysis results are summarised in Table 1 for $[Mo_3S_4]^{4+}$ and Table 2 for $[Mo_3S_{13}]^{2-}$, which provide a direct comparison between the experimentally observed Raman frequencies, the corresponding DFT-calculated frequencies, and previously reported values from the literature.[39–46] These tables also include vibrational assignments for each mode, providing a comprehensive reference dataset for the $[Mo_3S_4]^{4+}$ and $[Mo_3S_{13}]^{2-}$ cluster families. While the overall agreement between the experimental and DFT-predicted Raman spectra is strong, some discrepancies in peak positions and relative intensities are observed. These are expected and can be attributed to several factors. Frequency deviations between experiment and calculated are usually within ±10 $cm^{-1}$ or less. Such differences arise from the harmonic approximation used in DFT calculations, neglecting real systems' anharmonic effects.[64] Additionally, environmental influences such as crystal packing, intermolecular interactions, or solvent residues in the experimental samples can shift vibrational frequencies relative to the isolated-molecule model used in calculations.[65,66] Differences in Raman intensity between the experimentally obtained and calculated spectra are also more pronounced and commonly observed in molecular vibrational studies.[67] While DFT estimates Raman activity based on changes in polarisability, the experimental intensity depends on several additional factors: excitation wavelength, resonance enhancement, local dielectric environment, and instrumental response.[68] Moreover, subtle changes in the geometry or symmetry of a cluster, especially in the $[Mo_3S_{13}]^{2-}$ case, can lead to significant intensity redistribution among closely spaced modes. In this context, the presence of overlapping bands and the relative weakness of specific low- or high-frequency vibrations may lead to the underrepresentation or absence of certain theoretically predicted peaks in the experimental spectra. Despite these limitations, a very good agreement between the experimentally obtained and DFT-calculated Raman spectra is observed in Fig. 3, providing a solid reference Raman framework for future studies.

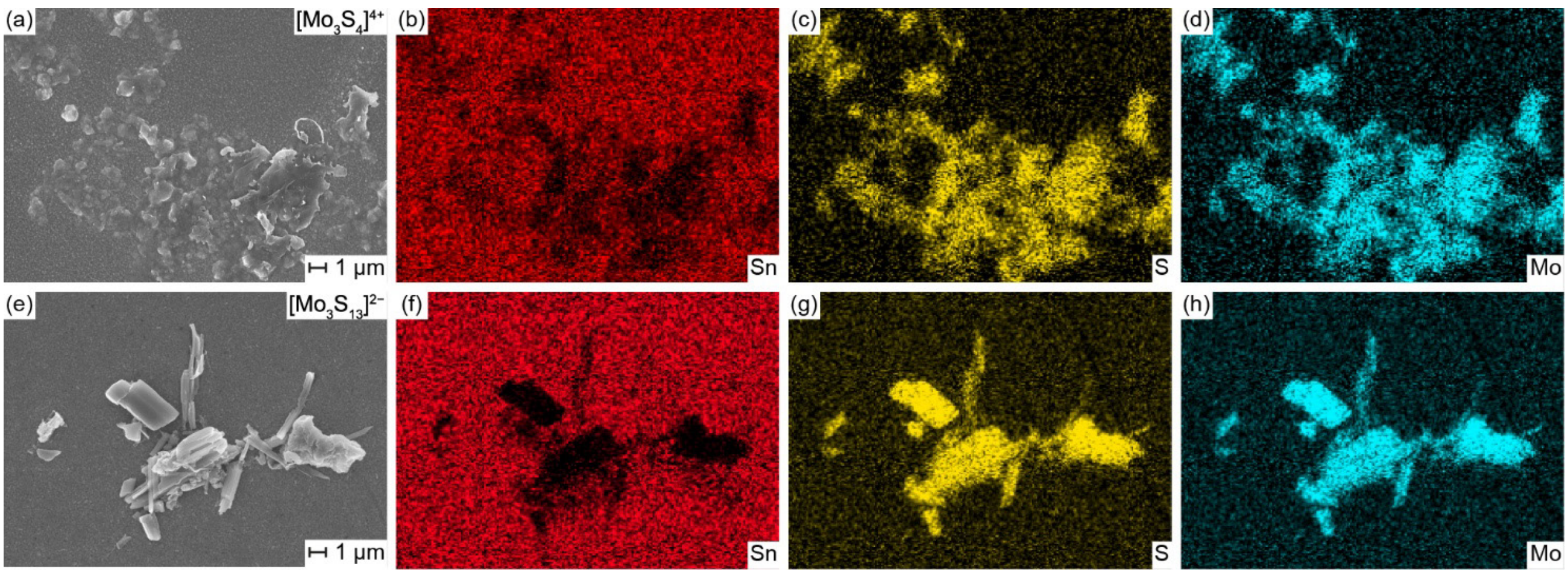

**Figure 2.** SEM spectra and EDX mapping of Sn, S and Mo for (a-d) $[Mo_3S_4]^{4+}$, (b-h) $[Mo_3S_{13}]^{2-}$.

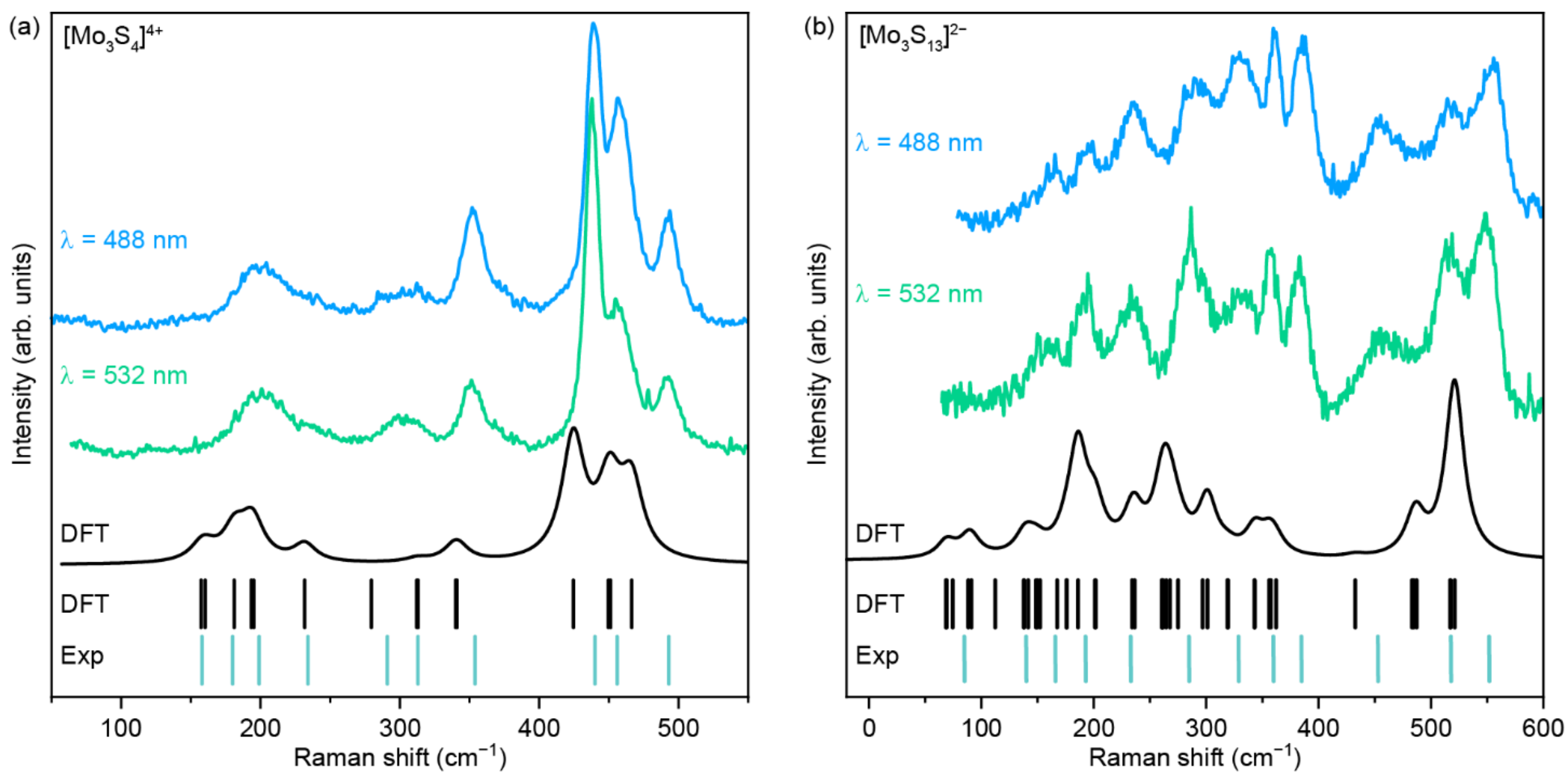

**Figure 3.** Raman spectra of (a) $[Mo_3S_4]^{4+}$ and (b) $[Mo_3S_{13}]^{2-}$ measured using 488 nm and 532 nm laser excitation and compared with DFT-calculated Raman spectra. Vertical lines under the spectrum show a comparison between the Raman peak positions obtained experimentally by the deconvolution (labelled "Exp") and from the lattice dynamics calculations based on DFT (labelled "DFT").

## *2.3 Vibrational Nature of Raman Modes in $[Mo_3S_4]^{4+}$*

**Table 1.** Frequencies ν (in $cm^{-1}$) of Raman peaks for $[Mo_3S_4]^{4+}$ obtained from Lorentzian fitting of experimental spectra and proposed vibrational mode assignments, compared with DFT-calculated values and literature references.

| This work | | | References | |
|---|---|---|---|---|
| Experimental ν ($cm^{-1}$) | DFT ν ($cm^{-1}$) | Assignment | ν ($cm^{-1}$) | Reference |
| 158 | 157, 160 | In-plane bending, Mo–S–Mo bridge bend | 141–160 | Sukhanova et al. (2023)[39] |
| 180 | 181 | Mo–S stretch | 180–210 | Sukhanova et al. (2023)[39] |
| 199 | 193, 195 | Mo–Mo + S–bridge coupled mode | | |
| 234 | 231 | Symmetric Mo–S–Mo stretch | 230–240 | Ohki et al. (2019)[40] |
| 291 | 280 | Asymmetric Mo–S stretch | 284 | Sukhanova et al. (2023)[39] |
| 313 | 312, 313 | Mixed bending/stretching | ~310 | Tran et al. (2016),[41] Deng et al. (2016)[42] |
| 354 | 340, 341 | Asymmetric S stretch | ~350 | Ohki et al. (2019)[40] |
| 440 | 424 | S-atom vibration (terminal/apical) | 440 | Deng et al. (2016)[42] |
| 456 | 449, 451 | S–S or Mo–S asymmetric bending | | |
| 493 | 466 | Terminal S stretch or Mo–S–S | 493 | Sukhanova et al. (2023)[39] |

In order to better understand the structural dynamics of the $[Mo_3S_4]^{4+}$ and $[Mo_3S_{13}]^{2-}$ clusters, their vibrational patterns were investigated, and the atomic displacements associated with key Raman-active modes were analysed. This enables the linking of the specific spectral features to distinct types of motion within the Mo–S framework. The vibrational modes of the $[Mo_3S_4]^{4+}$ cluster, presented in Fig. S1, reveal a particular interplay between metal–metal and metal–ligand dynamics that reflect the unique structure of the cluster. These figures show the atomic displacement patterns of all vibrational modes of $[Mo_3S_4]^{4+}$. Low-frequency modes (e.g., ~157–195 $cm^{-1}$) are predominantly characterised by concerted motions of Mo atoms within the $Mo_3$ triangle. These motions range from symmetric breathing-like displacements (e.g., 157 $cm^{-1}$), where all three Mo atoms move in and out synchronously relative to the triangle centre, to shearing or rocking-type modes (e.g., 195 $cm^{-1}$) involving relative lateral displacements of Mo atoms, which distort the triangle without changing its area. Such metal-core-dominated vibrations are generally indicative of strong Mo–Mo bonding interactions, which are central to the cluster's structural integrity and electronic properties. At intermediate frequencies (~230–340 $cm^{-1}$), the vibrations begin to involve more complex Mo–S stretching and bending, particularly with contributions from the apical sulphur atoms that cap the $Mo_3$ base. These modes typically display mixed character, where both Mo–Mo and Mo–S displacements are non-negligible. For instance, the mode near 231 $cm^{-1}$ shows cooperative displacement of all three Mo atoms accompanied by out-of-phase S atom motion, resembling umbrella-like deformations of the $Mo_3S_3$ unit. In contrast, the higher modes in this region feature localised stretching of Mo–S bonds with minor Mo participation. At higher frequencies (~424–466 $cm^{-1}$), the Raman-active modes are dominated by strong symmetric and asymmetric Mo–S stretching vibrations. These modes typically involve terminal sulphurs bonded to a single Mo centre and represent relatively localised vibrations compared to the lower-frequency delocalised modes. Notably, the 466 $cm^{-1}$ mode exhibits a pronounced symmetric Mo–S stretch consistent with literature reports of terminal Mo–S bond vibrations in related molybdenum–sulphur systems. Comparison with literature, particularly the works by Sukhanova et al. (2023)[39] and Ohki et al. (2019)[40] confirms the overall trends observed in our calculations. The dominant low-frequency modes reported experimentally (e.g., near 180–200 $cm^{-1}$) have been attributed to $Mo_3$ breathing or twisting motions, which aligns well with our assignments. Similarly, the high-frequency features in the ~450–470 cm$^{-1}$ range are assigned in the literature to terminal Mo–S stretches, consistent with our calculated eigenvectors. Though less discussed experimentally due to their lower Raman intensity, the modes in the intermediate range (300–350 cm$^{-1}$) are predicted to involve both bending and stretching with partial Mo character, corroborated by the literature.[42,46] Together, the calculated modes for $[Mo_3S_4]^{4+}$ support a coherent vibrational profile that reflects both the core rigidity of the $Mo_3$ triangle and the flexibility of the surrounding sulphur environment.

For identification in the Raman spectrum, $[Mo_3S_4]^{4+}$ exhibits prominent peaks at 158–200 cm$^{-1}$ from Mo–S–Mo bridge bending and $Mo_3$ core breathing or shearing modes, ~234 cm$^{-1}$ from symmetric Mo–S–Mo stretching, 291–354 cm$^{-1}$ from asymmetric Mo–S stretching and mixed bending/stretching vibrations, and 424–466 cm$^{-1}$from symmetric and asymmetric stretches of terminal sulphur atoms. The combination of the strong mid-frequency mode near 234 cm$^{-1}$ with the high-frequency Mo–S stretching bands enables straightforward recognition of $[Mo_3S_4]^{4+}$ in complex or supported materials.

## *2.4 Vibrational Mode Description of $[Mo_3S_{13}]^{2-}$*

The vibrational spectrum of the $[Mo_3S_{13}]^{2-}$ cluster spans a broad range from below 100 cm$^{-1}$ up to ~520 cm$^{-1}$, reflecting the presence of both heavy Mo–Mo/Mo–S framework motions and lighter terminal S–S vibrations. Fig. S2 in the Supporting Information shows the atomic displacement patterns of all vibrational modes of $[Mo_3S_{13}]^{2-}$. The vibrational modes can be categorised into distinct groups based on frequency ranges and atomic character. At the low end of the spectrum (below ~150 cm$^{-1}$), the modes are dominated by collective torsional and wagging motions involving the rigid $Mo_3$ triangle and its bridging S ligands (e.g., 68, 74 cm$^{-1}$). These modes often involve symmetric and antisymmetric twisting of the $Mo_3$ core, as well as out-of-plane librations of the bridging S atoms that connect Mo centres. Some of these low-frequency modes also include the concerted motion of the terminal disulphide ($S_2$) units, which oscillate relative to the $Mo_3$ frame with minimal internal deformation (e.g., 88 cm$^{-1}$).

**Table 2.** Frequencies ν (in cm$^{-1}$) of Raman peaks for $[Mo_3S_{13}]^{2-}$ obtained from Lorentzian fitting of experimental spectra and proposed vibrational mode assignments, compared with DFT-calculated values and literature references.

| This work | | | References | |
|---|---|---|---|---|
| Experimental ν (cm$^{-1}$) | DFT ν (cm$^{-1}$) | Assignment | ν (cm$^{-1}$) | Reference |
| 85 | 88 | Mo–S cage deformation | 85–90 | Xi et al. (2022),[43] Fedin et al. (1989)[44] |
| 140 | 137, 138, 142 | Internal Mo–S deformation | ~140 | Tran et al. (2016),[41] Fedin et al. (1989)[44] |
| 166 | 167 | In-plane Mo–S bend | 160–170 | Sukhanova et al. (2023)[39] |
| 193 | 186 | Mo–Mo stretch | 193 | Xi et al. (2022),[43] Müller et al. (1991)[45] |
| 233 | 234, 236 | Mo–S–Mo bridge motion | ~230 | Ohki et al. (2019)[40] |
| 285 | 275, 297 | Asymmetric Mo–S vibration | 284–290 | Tran et al. (2016)[41] |
| 329 | 319, 320 | Mixed stretching/bending | 325–330 | Tran et al. (2016)[41] |
| 360 | 356, 358 | S–S bridge bend | | |
| 385 | 362 | Terminal S wagging | 384 | Xi et al. (2022)[43] |
| 453 | 432 | S–S symmetric stretch | 450–460 | Müller et al. (1991)[45] |
| 518 | 517, 518 | Terminal S–S bond vibration | 518 | Xi et al. (2022),[43] Fedin et al. (1989),[44] Weber et al. (1995)[46] |

More complex Mo–S skeletal deformations are observed in the intermediate range between 150 and 300 cm$^{-1}$. These include in-plane scissoring and rocking of Mo–S–S units (e.g., 213, 275 cm$^{-1}$), as well as asymmetric flexing modes of the $S_2$ bridges. Several modes in this region exhibit a characteristic mixture of metal-ligand motion and internal S–S stretch or bend components, reflecting the hybridised nature of bonding within the cluster. Notably, Mo–S bond stretch and bend vibrations become more prominent here (e.g., 296 cm$^{-1}$), with some differentiation between inner and outer Mo–S bonds depending on the local geometry. Above 300 cm$^{-1}$, the vibrations increasingly localise on the sulphur atoms, especially on the terminal $S_2$ ligands. This includes internal S–S stretching vibrations, where each disulphide unit oscillates nearly independently (e.g., 343, 357 cm$^{-1}$). These modes exhibit significant Raman activity due to the substantial polarisability changes associated with S–S bond deformation. Near the upper end of the spectrum (above ~480 cm$^{-1}$), the most localised and high-energy S–S stretching modes are found (e.g., 484, 487 cm$^{-1}$). These involve almost pure terminal S–S stretches with minimal coupling to the Mo framework and often appear as sharp peaks in experimental Raman spectra.

Across the vibrational spectrum, symmetry trends can be observed: low-frequency modes tend to preserve the overall molecular symmetry, while high-frequency modes increasingly break symmetry due to the localised nature of the atomic displacements, particularly within terminal ligands. In many cases, the threefold symmetry of the cluster leads to near-degenerate mode pairs or subtle splittings due to structural asymmetry in the relaxed geometry. The rich coupling between Mo–S and S–S motions, particularly in the mid-frequency region, contributes to the vibrational complexity and underscores the mixed covalent/ionic character of bonding in $[Mo_3S_{13}]^{2-}$. These insights are valuable for interpreting experimental Raman and IR spectra, particularly in identifying fingerprint regions associated with $S_2$ ligands and Mo–S connectivity.

Several of the vibrational features identified here are consistent with previous Raman studies on $[Mo_3S_{13}]^{2-}$ and related clusters. High-frequency bands around 450–520 cm$^{-1}$, attributed to terminal S–S stretching, were also reported by Xi et al. (2019)[43] for $[Mo_3S_{13}]^{2-}$ films on FTO and by Tran et al. (2016)[41] in electrodeposited $MoS_x$ films derived from $[Mo_3S_{13}]^{2-}$. Modes in the 200–300 cm$^{-1}$ range, linked to Mo–S bending and scissoring, have also been observed in $[Mo_3S_{13}]^{2-}$ based systems, as reported by Ohki et al. (2019).[40] The analysis expands on these observations, providing a detailed assignment of individual vibrational patterns.

In this context, $[Mo_3S_{13}]^{2-}$ can be readily identified by its broader Raman signature, which includes bands at 85–140 cm$^{-1}$ from $Mo_3$ framework deformations and bridging sulphur twisting, 166–233 cm$^{-1}$ from in-plane Mo–S bending and Mo–S–Mo bridge motion, and 285–360 cm$^{-1}$ from asymmetric Mo–S vibrations and S–S bridge bending. A mode near 385 cm$^{-1}$ corresponds to terminal sulphur wagging, while sharp peaks at ~453 and ~518 cm$^{-1}$ are characteristic of symmetric stretching of terminal disulphide ($S_2^{2-}$) ligands. These intense S–S stretching modes, in combination with the lower-frequency Mo–S features, form a unique Raman fingerprint that enables unambiguous detection of $[Mo_3S_{13}]^{2-}$ even at low surface coverage or within mixed-phase systems.

## *2.5 Infrared Spectra and Complementary Vibrational Analysis*

In addition to Raman spectroscopy, the infrared (IR) spectra of the $[Mo_3S_4]^{4+}$ and $[Mo_3S_{13}]^{2-}$ clusters were also calculated and measured to explore their vibrational fingerprints further and investigate the complementary activity of vibrational modes. The simulated IR spectra, shown in Fig. 4, were derived from DFT calculations performed without symmetry constraints, ensuring that all IR-active modes are represented regardless of symmetry selection

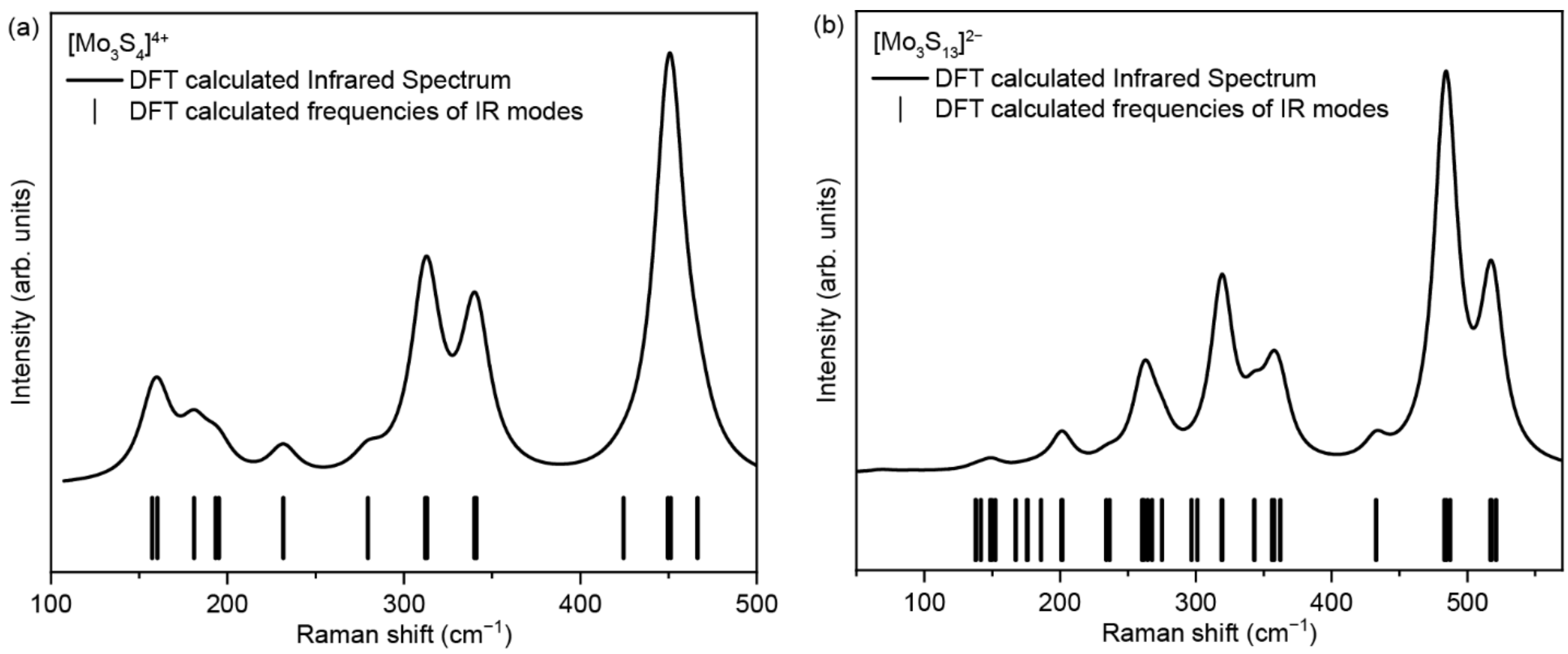

**Figure 4.** Calculated infrared (IR) spectra of the $[Mo_3S_4]^{4+}$ and $[Mo_3S_{13}]^{2-}$ clusters. (a) $[Mo_3S_4]^{4+}$ exhibits IR-active modes primarily in the 250–500 cm$^{-1}$ range, corresponding to Mo–S stretching and bending vibrations. (b) $[Mo_3S_{13}]^{2-}$ shows a broader and more complex IR spectrum, with intense features in the 150–520 cm$^{-1}$ region, including strong terminal S–S stretching bands above 450 cm$^{-1}$.

rules. The measured IR spectrum is given in Figure S4 and S5 in the Supporting Information.

Direct overlap of the measured and DFT-calculated IR spectra was not possible because the experimental data were collected only within the mid-IR transmission window, which excludes much of the low-frequency (< 400 cm$^{-1}$) region where the majority of cluster-related modes are predicted. Furthermore, the experimental spectra are dominated by intense, broad absorptions from O–H stretching and bending vibrations of coordinated and lattice water, as well as possible N–H modes from ammonium counterions. These strong high-frequency features obscure the weaker Mo–S and S–S vibrations in the accessible range, making a direct mode-by-mode comparison with the full DFT-predicted spectrum impractical. As a result, we present the experimental IR data separately and use the calculated spectra to guide mode assignments in the low-frequency region that could not be measured.

As shown in Fig. 4a, for $[Mo_3S_4]^{4+}$, the IR spectrum is dominated by bands in the mid-frequency region between 250 and 500 cm$^{-1}$, corresponding primarily to Mo–S stretching and Mo–S–Mo bending vibrations. These features largely complement the Raman-active modes. The absence of a high density of IR-active modes in the low-frequency region suggests that these motions, primarily involving Mo–Mo deformation, are only weakly coupled to changes in dipole moment, as expected for metal-metal breathing modes.

In contrast, in Fig. 4b, the $[Mo_3S_{13}]^{2-}$ cluster displays a significantly richer IR spectrum, with multiple intense bands extending from 150 to over 500 cm$^{-1}$. These include strong IR-active vibrations associated with bridging and terminal S–S stretching, particularly in the 450–520 cm$^{-1}$ region. These modes are prominent in IR due to their strong dipole moment variation, making them particularly useful for confirming the presence and integrity of disulphide ligands. Additionally, the mid-frequency IR-active modes around 250–350 cm$^{-1}$ provide complementary information on Mo–S–S bending and scissoring vibrations. These calculations highlight the value of IR analysis as a complementary tool for full vibrational mode identification in complex Mo–S clusters. This provides a reference for future experimental IR studies to refine cluster characterisation.

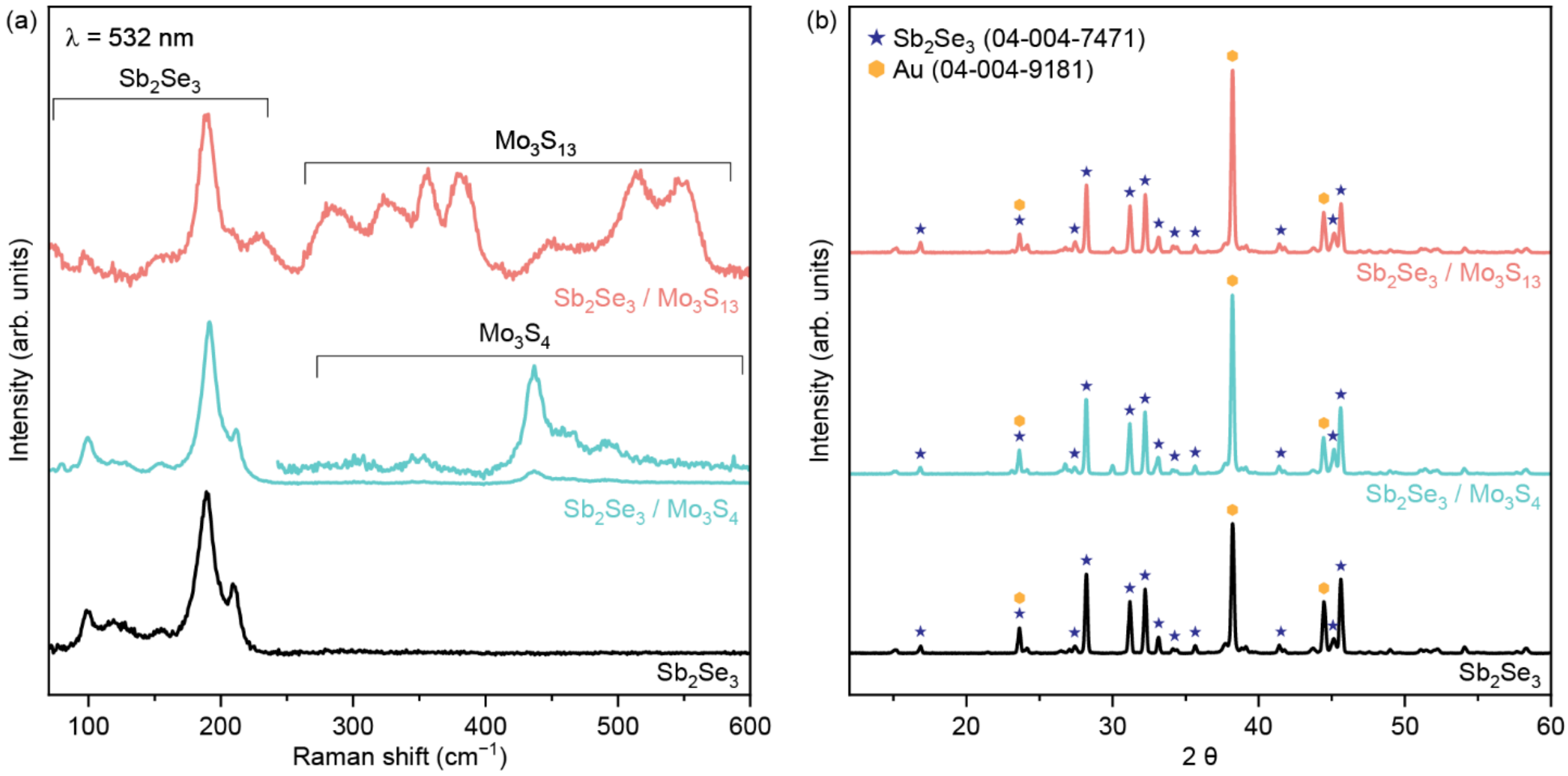

**Figure 5.** (a) Raman spectra of $Sb_2Se_3$ thin films before and after deposition of $[Mo_3S_4]^{4+}$ and $[Mo_3S_{13}]^{2-}$ clusters. The reference $Sb_2Se_3$ film shows its characteristic Raman features, while the addition of $[Mo_3S_4]^{4+}$ and $[Mo_3S_{13}]^{2-}$ introduces distinct vibrational bands corresponding to Mo–S stretching $[Mo_3S_4]^{4+}$ and S–S disulphide stretching $[Mo_3S_{13}]^{2-}$, enabling precise identification of each cluster. (b) X-ray diffraction (XRD) patterns of the same three samples. No structural changes are observed upon cluster deposition, illustrating the limited sensitivity of XRD compared to Raman spectroscopy for detecting surface-bound molecular species.

## *2.6 Raman Spectroscopy as a Structural Fingerprinting Tool*

Fig. 5 highlights the practical advantage of Raman spectroscopy over X-ray diffraction (XRD) in distinguishing Mo-based clusters deposited on complex substrates. In this example, $[Mo_3S_4]^{4+}$ and $[Mo_3S_{13}]^{2-}$ clusters were deposited on thin films of $Sb_2Se_3$, a widely studied light-absorbing material for water-splitting photocathodes.[69–72] The goal was to verify whether the two cluster types could be unambiguously detected when supported on the same photoabsorber substrate. Fig. 5a presents Raman spectra from three samples: bare $Sb_2Se_3$, $Sb_2Se_3$ coated with $[Mo_3S_4]^{4+}$, and $Sb_2Se_3$ coated with $[Mo_3S_{13}]^{2-}$. The spectrum of the bare $Sb_2Se_3$ film shows a typical Raman signature of $Sb_2Se_3$, characterised by features in the low- to mid-frequency region (< 250 $cm^{-1}$). Additional peaks emerge upon deposition of the $[Mo_3S_4]^{4+}$ cluster, including characteristic bands above 300 $cm^{-1}$, consistent with Mo–S stretching identified earlier.

In the case of $[Mo_3S_{13}]^{2-}$ deposition, several new peaks appear in the high-frequency region, most in the 250–520 $cm^{-1}$ range, corresponding to vibrational pattern of $[Mo_3S_{13}]^{2-}$ presented earlier. These features clearly differentiate between the two cluster types and demonstrate Raman's ability to resolve their unique vibrational fingerprints, even when deposited as thin surface layers. In contrast, Fig. 5b presents the corresponding XRD patterns for the same three samples. All patterns exhibit reflections characteristic of the crystalline $Sb_2Se_3$ phase, but no additional peaks or shifts are observed upon deposition of either cluster. This is likely due to the small size, low crystallinity, or surface-dispersed nature of the Mo clusters, which are often invisible to conventional diffraction techniques. This comparison showcases the superior sensitivity of Raman spectroscopy to short-range structural order and local bonding environments over XRD. In contexts where the clusters are amorphous, disordered, or supported in sub-monolayer quantities, Raman measurements provide clear vibrational signatures that allow unambiguous identification. This highlights its value for fundamental characterisation and in-situ or operando studies of functional catalysts and photoelectrodes.

## 3. Conclusion

In this work, a comprehensive vibrational study of the $[Mo_3S_4]^{4+}$ and $[Mo_3S_{13}]^{2-}$ clusters were carried out through a combination of experimental Raman spectroscopy and first-principles calculations. Raman spectra acquired with 488 nm and 532 nm excitation were compared to DFT-calculated vibrational modes, enabling full mode assignment across the entire frequency range. For $[Mo_3S_4]^{4+}$, characteristic Raman bands were observed in the 160–230 $cm^{-1}$ region, associated with $Mo_3$ core breathing and Mo–S–Mo bending modes, as well as in the 300–460 $cm^{-1}$ range, dominated by Mo–S stretching vibrations. For $[Mo_3S_{13}]^{2-}$, a broader vibrational signature was identified, with distinct Raman bands appearing from 70 $cm^{-1}$ up to 520 $cm^{-1}$, including strong S–S stretching modes around 453 and 518 $cm^{-1}$—serving as reliable fingerprints for the disulphide ligands. Complementary calculations of the IR spectra provided additional information about the vibrational properties and mode of activity of these clusters. The combined experimental and theoretical analysis presented here establishes a complete vibrational reference for $[Mo_3S_4]^{4+}$ and $[Mo_3S_{13}]^{2-}$, offering valuable tools for their structural identification. Moreover, it is demonstrated that Raman spectroscopy enables unambiguous detection of these clusters even in complex environments, such as when supported on $Sb_2Se_3$ photocathodes, where XRD analysis shows no clear indication of their presence. This work establishes a robust framework for spectroscopic identification and structural analysis of $Mo_3$-based clusters and serves as a foundation for future studies in catalysis and energy conversion applications.

## Author Contributions

M.D. and P.A. conceptualised the project, curated and analysed data, and wrote the manuscript. J.B. synthesised the catalyst and aided in the manuscript's visualisation aspects, reviewing and editing. A.L.A. did DFT. M.D. acquired the Raman Data. M.D. and S.D.T. supervised the project.

## Conflicts of Interest

There are no conflicts to declare.

## Acknowledgements

P.A. acknowledges financial support from the UZH Entrepreneur Fellowship of the University of Zürich, reference no. [SUSEF24-001]. P.A. and S.D.T. would like to thank the University of Zürich and the Swiss National Science Foundation (Project #214810) for support.

# Raman and IR Signatures of [$Mo_3S_4$]$^{4+}$ and [$Mo_3S_{13}$]$^{2-}$ Molybdenum Sulphide Molecular Catalysts for Solar Hydrogen Evolution

**Pardis Adams[1], Jan Bühler[1], Angel Labordet Alvarez[2,3], D. David Tilley[1], and Mirjana Dimitrievska[2,*]**

[1] Department of Chemistry, University of Zurich, Zurich, Switzerland.
[2] Nanomaterials Spectroscopy and Imaging, Transport at Nanoscale Interfaces, Swiss Federal Laboratories for Materials Science and Technology (Empa), Dubendorf, Switzerland.
[3] Department of Physics and Swiss Nanoscience Institute, University of Basel, Basel, Switzerland

*Author to whom any correspondence should be addressed.

E-mail: Mirjana.Dimitrievska@empa.ch

## Table of Contents

## General Information

All chemicals were of reagent grade or higher, obtained from commercial sources and used without further purification. Solvents for reactions were of p.a. grade; $H_2O$ was ultrapure from a *Milli-Q® Direct 8* water purification system.

**FT-IR** spectra were recorded on a *SpectrumTwo* FT-IR Spectrometer (*Perkin–Elmer*); samples were applied as KBr pellets. **High-resolution electrospray mass spectra (HR-ESI-MS)** were recorded on a *timsTOF Pro TIMS-QTOF-MS* instrument (*Bruker Daltonics GmbH*, Bremen, Germany). The samples were dissolved in MeOH at a ca. 50 µg $mL^{-1}$ concentration and analysed via continuous flow injection (2 µL $min^{-1}$). The mass spectrometer was operated in the positive or negative electrospray ionization mode at 4'000 V (-4'000 V) capillary voltage and −500 V (500 V) endplate offset with an $N_2$ nebulizer pressure of 0.4 bar and a dry gas flow of 4 L $min^{-1}$ at 180 °C. Mass spectra were acquired in a mass range from m/z 50 to 2'000 at ca. 20'000 resolution (m/z 622) and at 1.0 Hz rate. The mass analyser was calibrated between m/z 118 and 2'721 using an *Agilent ESI-L* low-concentration tuning mix solution (*Agilent*, USA) at a resolution of 20,000, giving a mass accuracy below 2 ppm. All solvents used were purchased in the best LC-MS quality. **UV-Vis** spectra were recorded on a S*himadzu UV-3600 Plus* spectrophotometer.

## Syntheses

**$[Mo_3S_4(H_2O)_9]Cl_4$** (**1**, prepared according to a published procedure[1,2])

Caution: Toxic $H_2S$ is evolved during the reaction. It is recommended to connect gas-washing bottles containing bleach to the set-up.

A solution of $(NH_4)_2[MoS_4]$ (1.00 g, 3.84 mmol) in $H_2O$ (35 mL) was treated alternatingly with 1 mL portions of aqueous HCl (6 M, 15 mL) and aqueous $NaBH_4$ (2 M, 15 mL) under ambient conditions. The resulting brown mixture was heated to 90 °C for 22 h with an oil bath, while compressed air was continuously introduced via a glass syringe. Upon completion, the reaction mixture had turned dark green and was cooled with an ice bath before the resulting precipitate was filtered off and washed with 1 M HCl (10 mL). The combined filtrate was concentrated under reduced pressure to approximately 5 mL and purified by size-exclusion chromatography over a *Sephadex G-10* column (10 g, bloomed in 1 M HCl, eluted with 1 M HCl). A fraction containing the green $[Mo_3S_3O]^{4+}$ ($\lambda_{max}$ 605 nm) eluted first, followed by a darker green fraction ($\lambda_{max}$ 620 nm) corresponding to the product $[Mo_3S_4(H_2O)_9]^{4+}$. The fractions of $[Mo_3S_4(H_2O)_9]^{4+}$ were diluted with $H_2O$ (five times the original volume) and purified further by cation-exchange chromatography over a *DOWEX 50WX2* column (15 g, preconditioned with 2 M HCl, eluted with 2 M HCl). A light brown band containing $[Mo_2O_2S_2]^{2+}$ eluted first, followed by the dark green band of the product. Evaporation of the eluent under reduced pressure yielded $[Mo_3S_4(H_2O)_9]Cl_4$ as a dark green powder (362 mg, 1.28 mmol, 39%).

UV-Vis: $\lambda_{max}$ (1 M HCl)/nm 255 (ε/$dm^3$ $mol^{-1}$ $cm^{-1}$; 7'945), 371 (3'475), 620 (172).
FT-IR (KBr): $\nu_{max}/cm^{-1}$ 3390br, 3225s, 1622m, 1404m, 1195w, 960w, 847m, 806m, 648w, 570w, 549w.
HRMS (ESI): m/z calc. for $C_2H_6ClMo_3O_2S_4$ [M−9 $H_2O$+2 OMe+Cl]$^+$: 518.60958; found: 518.60871.

**$(NH_4)_2[Mo_3S_{13}]\cdot 2\ H_2O$** (**2**, prepared according to a published procedure[3])

A solution of $(NH_4)_2S_x$ (25 wt-%) was prepared by dissolving elemental sulfur (3.00 g) in a solution of $(NH_4)_2S$ (48 wt-%, 20 mL) in $H_2O$ (20 mL). Separately, $(NH_4)_2[Mo_7O_{24}]\cdot 4\ H_2O$ (1.02 g, 825 µmol) was dissolved in $H_2O$ (5 mL), and the freshly prepared $(NH_4)_2S_x$ solution (25 wt-%, 30 mL) was added. The reaction flask was covered with a watch glass and heated to 95 °C for 96 h with an oil bath and without stirring. Dark red crystals and pockets of elemental sulfur were formed and the solids were isolated by filtration. The filter cake was sequentially washed with $H_2O$ (3×10 mL), EtOH (3×10 mL), $CS_2$ (3×10 mL, until residual sulfur was fully removed), and $Et_2O$ (3×10 mL). The solid was air-dried to yield $(NH_4)_2[Mo_3S_{13}]\cdot 2\ H_2O$ as dark red crystals (1.32 g, 1.70 mmol, 88% yield).

UV-Vis: $\lambda_{max}$ (MeOH)/nm 267 (ε/$dm^3$ $mol^{-1}$ $cm^{-1}$; 38’978), 425 (4’432).
FT-IR (KBr): $\nu_{max}/cm^{-1}$ 3437br, 3081m, 2926m, 2781m, 1633w, 1566w, 1399s, 1385s, 544s, 506s, 459w.
HRMS (ESI): m/z calc. for $H_3Mo_3S_{13}$ [M+3 H]$^+$: 712.37607; found: 712.37391.

## Substrate preparation

*Pilkington's FTO TEC 15* substrates were cut into 1×2.5 cm pieces and cleaned by sonication in $H_2O$ containing alkaline detergent (Deconex 11 Universal), $H_2O$, acetone, and isopropyl alcohol. Subsequently, the substrates were dried with a stream of $N_2$ and underwent UV/ozone cleaning (30 min) to eliminate surface contaminants.

## Synthesis of $Sb_2Se_3$ Photocathodes

Prepared according to a published procedure.[4]
A *Safematic CCU-010* sputter coater was used to sequentially deposit a 10 nm layer of titanium (Ti) (acting as an adhesion layer) and a 150 nm layer of gold (Au) (serving as a hole-extracting electrode by creating an ohmic contact with the photoabsorber) onto the FTO substrates. The electrodeposition of antimony (Sb) metal was carried out with a three-electrode setup (FTO/Ti/Au stack, Pt wire, Ag/AgCl reference electrode). A solution of potassium antimony tartrate (15 mM) and tartaric acid (50 mM), adjusted to pH 1.3 by addition of $H_2SO_4$, served as the electrodeposition solution. A potential of −0.3 V *versus* Ag/AgCl was applied until a charge 1.4 C $cm^{-2}$ was passed. Subsequently, the Sb substrates underwent selenisation using a two-zone furnace. Selenium pellets were placed around the substrate, and the chamber was purged with argon. The temperature was increased by 15 °C $min^{-1}$ up to 350 °C before heating was continued for 40 min at 350 °C.

## Solution Treatment and Catalyst Deposition

Before catalyst deposition, the substrate (bare FTO or $Sb_2Se_3$ photocathode) was immersed into a solution of $(NH_4)_2S$ (10–12 wt%) for 5 s before rinsing with $H_2O$ and drying with a stream of $N_2$.
Catalyst deposition solutions (1 mM) of $[Mo_3S_4(H_2O)_9]Cl_4$ ($[Mo_3S_4]^{4+}$, 7.20 mg, 0.01 mmol) in aqueous HCl (1 M, 10 mL) and $(NH_4)_2[Mo_3S_{13}]\cdot 2\ H_2O$ ($[Mo_3S_{13}]^{2-}$, 7.77 mg, 0.01 mmol) in $H_2O$ (10 mL) were prepared by sonication for 30 min. Pre-treated samples were then placed in the catalyst solution for 12 h at ambient temperature. They were then rinsed with $H_2O$ from the backside and annealed at 120˚ C for 30 min.

## Morphology and Crystal Characterisation

Top-view **scanning electron microscopy (SEM)** and **energy-dispersive X-ray spectroscopy (EDX)** images were acquired using a *Zeiss Gemini 450* SEM device. **X-ray diffraction (XRD)** analysis utilized the *Rigaku Smartlab* diffractometer. Reference cards for $Sb_2Se_3$ (04-004-7471) and Au (04-004-9181) were obtained from the *Cambridge Crystallographic Data Centre* (CCDC) database.

## Raman Measurements

Raman measurements were performed using *a WITec Alpha 300 R* confocal Raman microscope operated in backscattering geometry. All samples were analysed using two excitation wavelengths: 488 and 532 nm. The laser beam was focused onto the sample using a high-numerical-aperture objective, yielding spot sizes of approximately 800 nm for 488 nm and 1 µm for 532 nm excitation. Laser power was carefully optimized to avoid thermal or photo-induced damage: power-dependent studies were conducted by gradually increasing the laser intensity while monitoring for any changes in spectral features such as peak position, broadening, or the appearance of new bands. The highest laser power that did not alter these characteristics was selected for final measurements. Backscattered Raman signals were collected using a 300 mm spectrometer equipped with an 1800 grooves/mm diffraction grating and a thermoelectrically cooled CCD detector. All spectra were calibrated against the Raman signal of a reference silicon sample to ensure spectral accuracy.

## Density Functional Theory (DFT)

All density functional theory (DFT) simulations were carried out using *ORCA 6.0.1*.[5] For each molecular species $[Mo_3S_4]^{4+}$ and $[Mo_3S_{13}]^{2-}$, gas-phase geometry optimizations were performed with the *Perdew–Burke–Ernzerhof* (PBE) exchange-correlation functional[6] in conjunction with the *def2-SVP* basis set.[7] The resolution of identity (RI) approximation was employed using the matching *def2/J* auxiliary basis set to improve computational efficiency, and the *TIGHTSCF* keyword was specified to ensure tighter self-consistent field (SCF) convergence. All geometry optimizations were done under default optimization criteria, with a maximum of 5000 optimization steps allowed where needed. Following optimization, vibrational frequency calculations were carried out within the harmonic approximation to confirm that each optimized structure was a true minimum (no imaginary frequencies) and to obtain infrared (IR) intensities. Raman activities were computed by enabling numerical frequency calculations (*via NumFreq*) and polarizability derivatives (*via %ELPROP POLAR 1*). Parallelization was performed using between 8 and 32 processes (specified by *%pal nprocs*) to reduce wall-clock time. The resulting IR and Raman spectra were derived directly from the computed vibrational frequencies, dipole derivatives, and polarizability derivatives, providing insights into the vibrational modes of these Mo–S clusters.

## Additional Figures

**Table S1** Atomic percentages of $[Mo_3S_4]^{4+}$ and $[Mo_3S_{13}]^{2-}$ on FTO, determined by energy-dispersive X-ray spectroscopy (EDX).

| | $[Mo_3S_4]^{4+}$ | $[Mo_3S_{13}]^{2-}$ |
|---|---|---|
| Element | Atomic % | Atomic % |
| O | 48.8 | 52.1 |
| Sn | 25.0 | 31.7 |
| C | 17.0 | 10.3 |
| S | 3.6 | 3.4 |
| Mo | 3.2 | 1.1 |

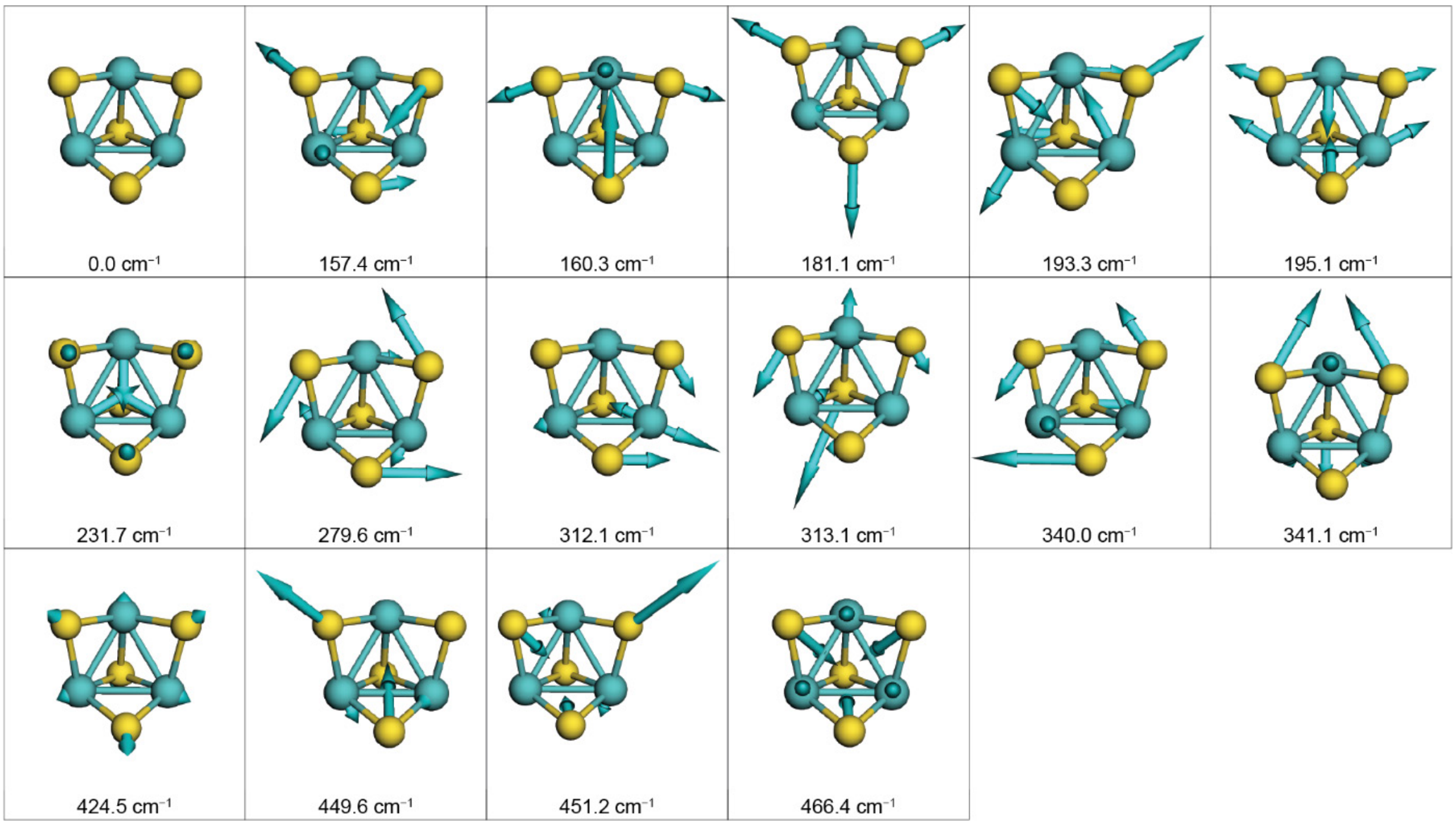

**Fig. S1** Calculated phonon displacements for Raman modes of $[Mo_3S_4]^{4+}$. Frequencies (in $cm^{-1}$) are listed under each image.

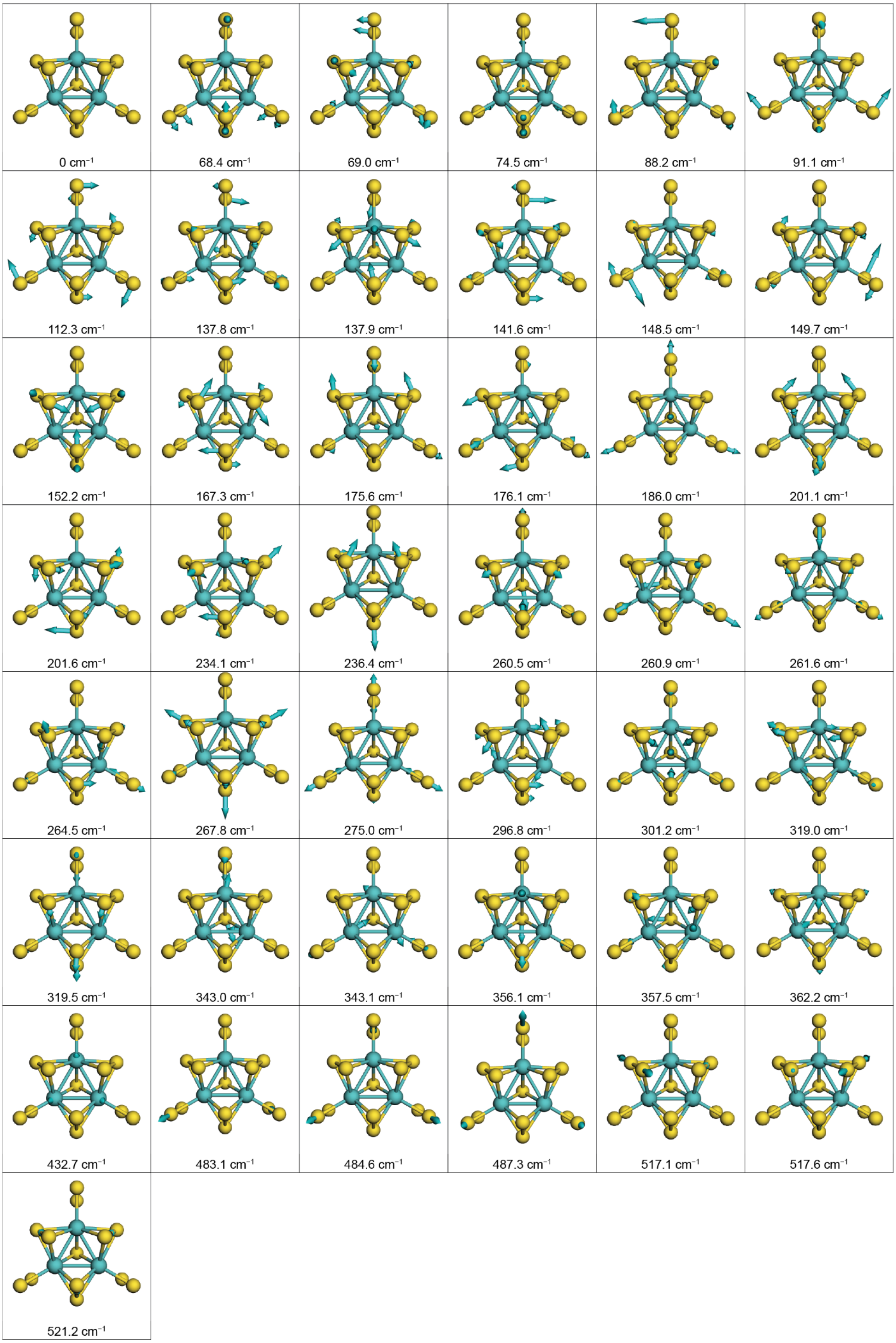

**Fig. S2** Calculated phonon displacements for Raman modes of $[Mo_3S_{13}]^{2-}$. Frequencies (in $cm^{-1}$) are listed under each image.

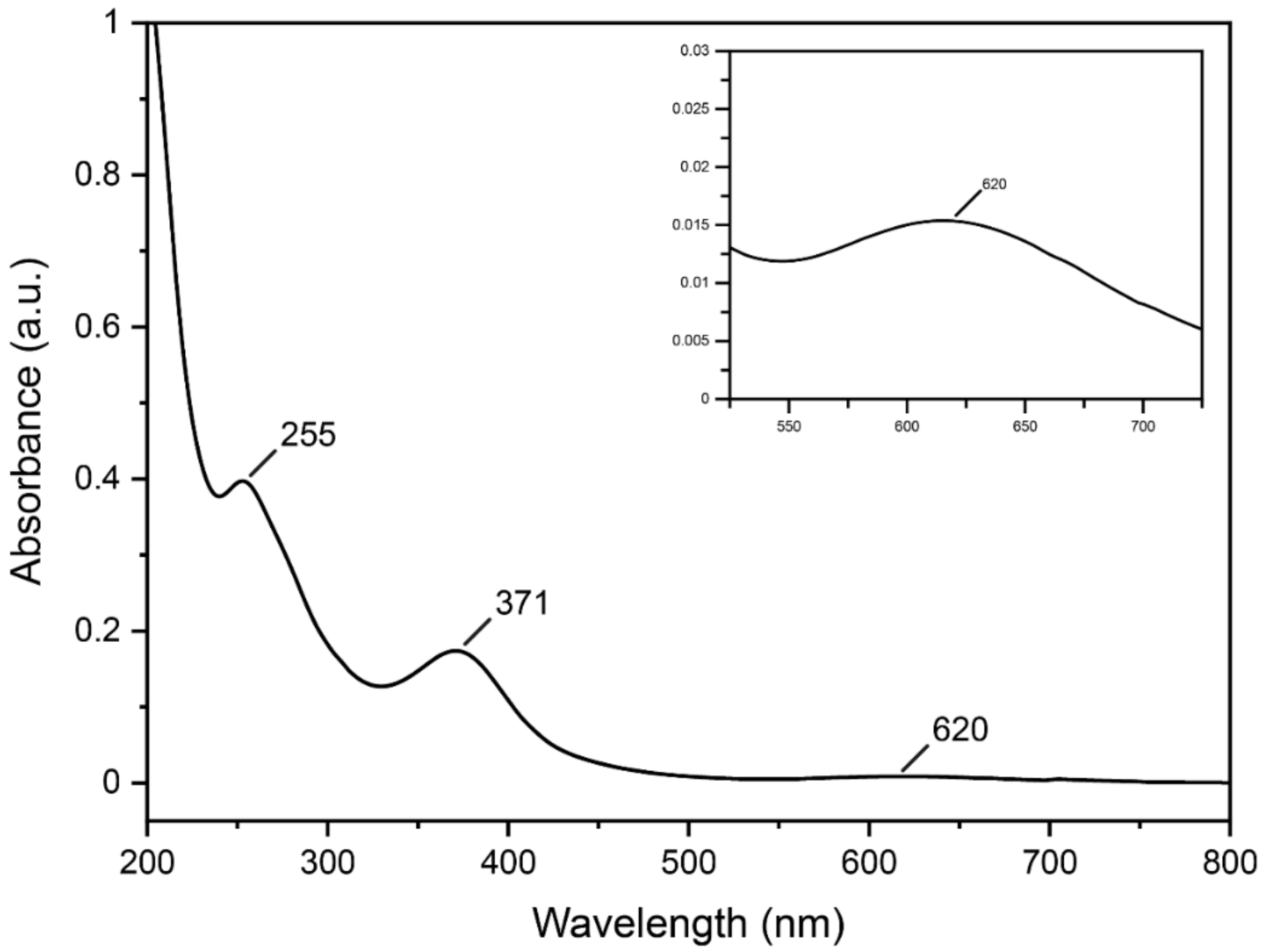

**Fig. S3** UV-Vis spectrum of $[Mo_3S_4(H_2O)_9]Cl_4$ (**1**), measured in 1 M HCl (50 mM).

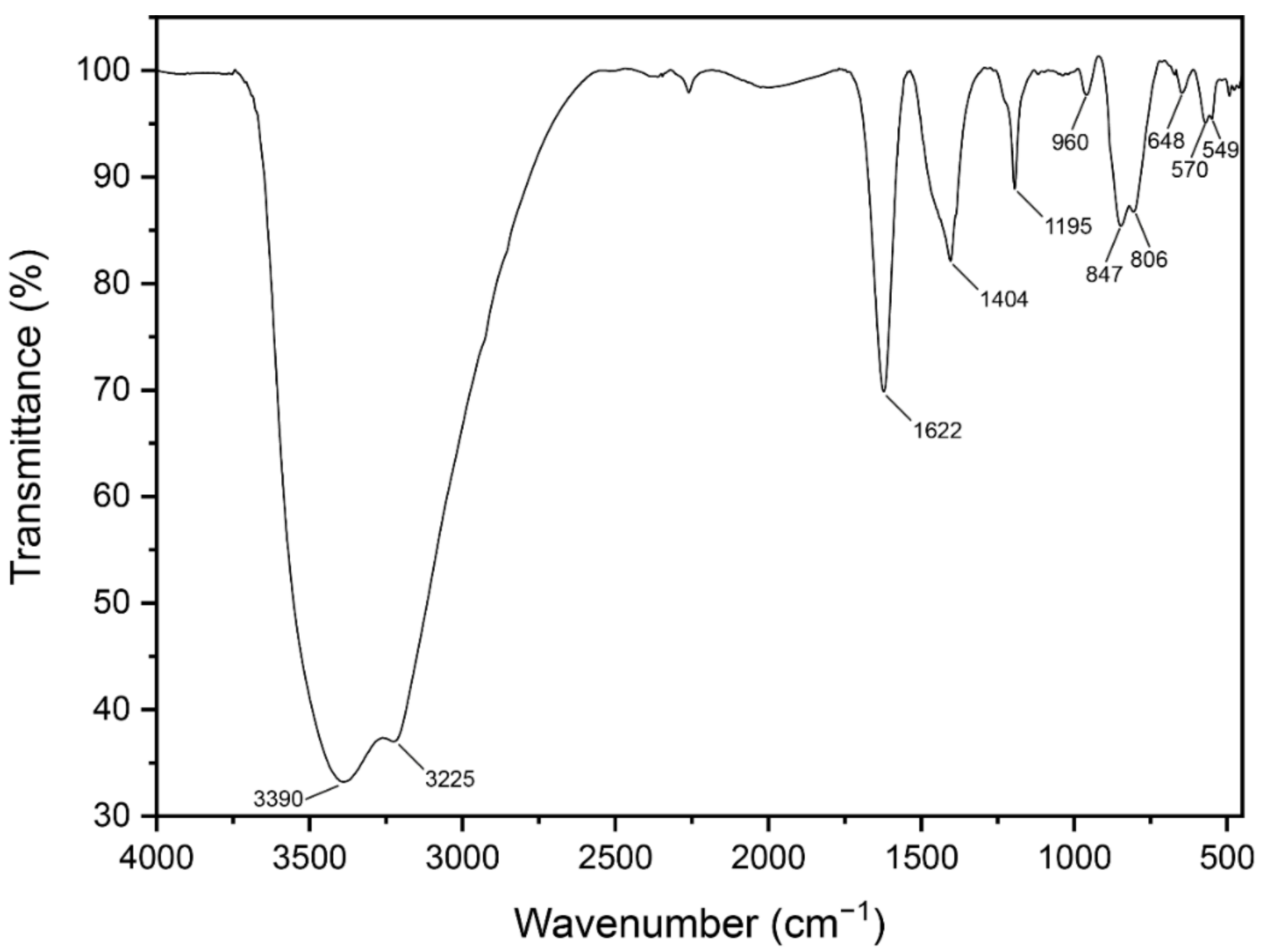

**Fig. S4** IR spectrum of $[Mo_3S_4(H_2O)_9]Cl_4$ (**1**), measured as KBr pellet.

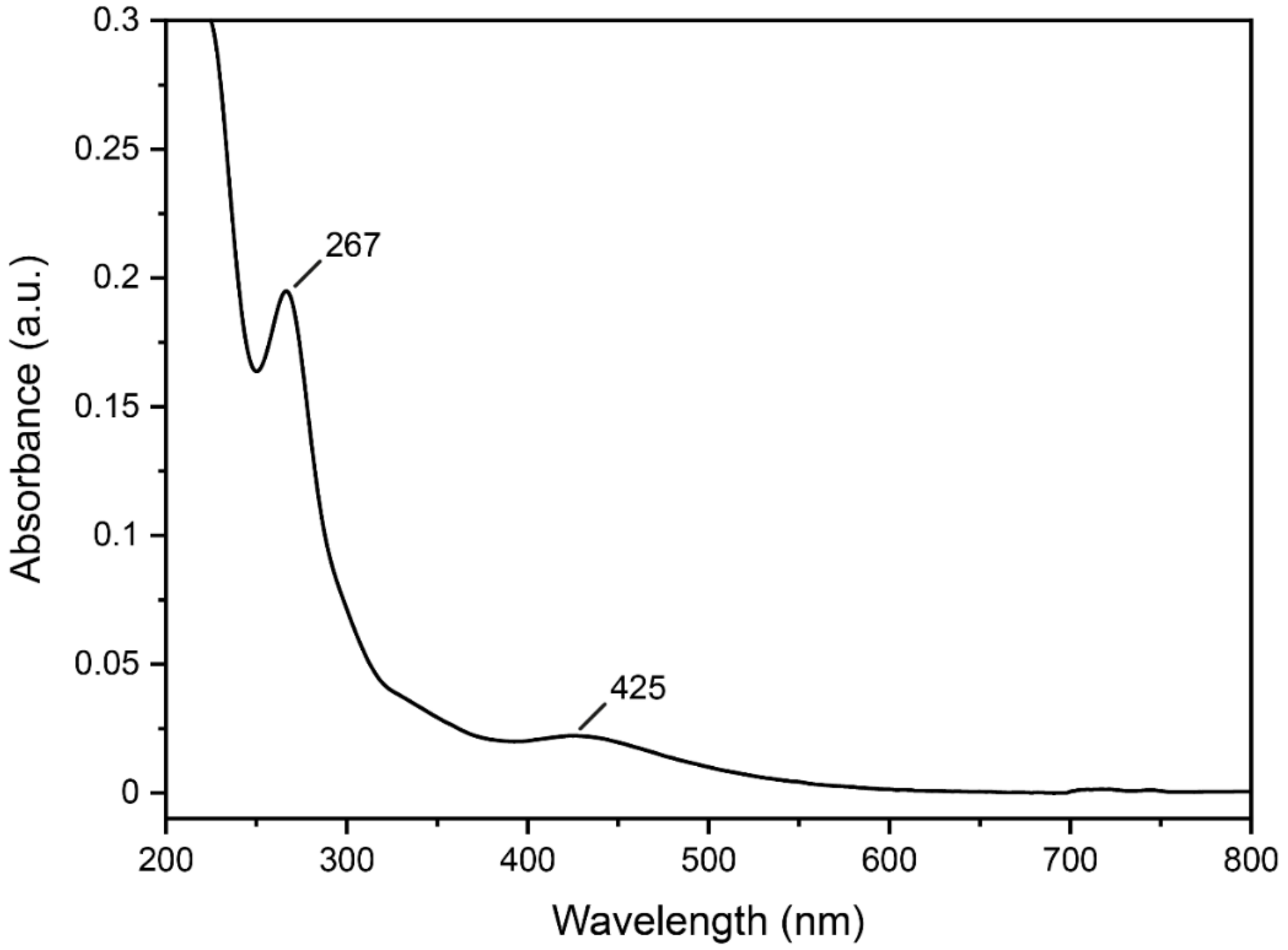

**Fig. S5** UV-Vis spectrum of $(NH_4)_2[Mo_3S_{13}]\cdot 2\ H_2O$ (**2**), measured in MeOH (5 mM).

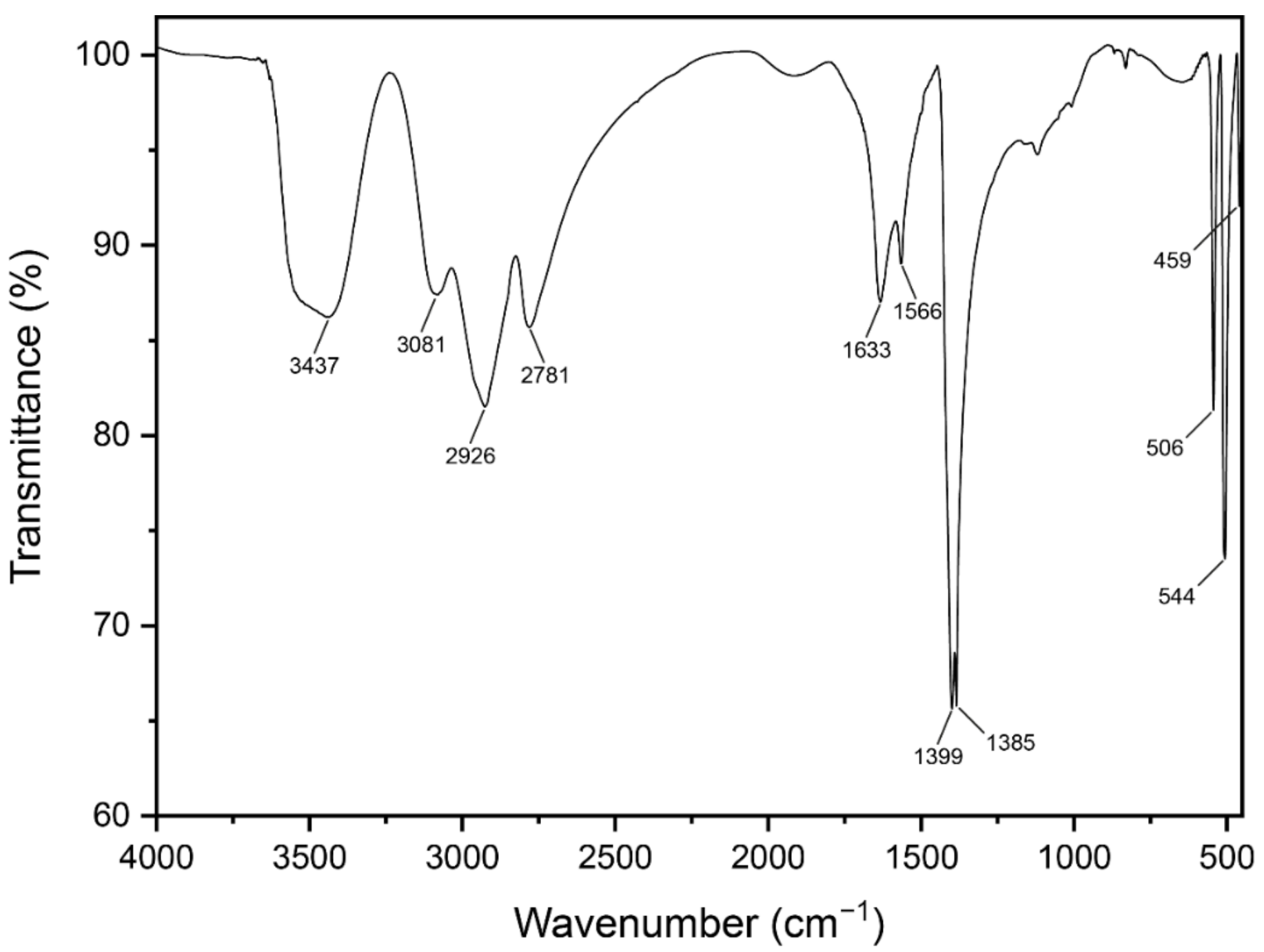

**Fig. S6** IR spectrum of $(NH_4)_2[Mo_3S_{13}]\cdot_2\ H_2O$ (**2**), measured as KBr pellet.